\documentstyle[proceedings,graphicx]{crckapb}

\renewcommand{\thepage}{}   

\def\vecn{\mbox{\boldmath $n$}}

\def\veck{\mbox{\boldmath $k$}}

\def\veccA{\mbox{\boldmath $c_A$}}

\def\vecr{\mbox{\boldmath $r$}}
\def\vecA{\mbox{\boldmath $A$}}
\def\vecC{\mbox{\boldmath $C$}}

\def\vecB{\mbox{\boldmath $B$}}
\def\vecU{\mbox{\boldmath $U$}}
\runningtitle{Acoustic tomography}  
\begin{opening}
  \title{Acoustic tomography of solar convective flows and structures}
  \author{A.G. Kosovichev}
  \institute{W.W. Hansen Experimental Physics Laboratory  \\
             Stanford University, CA 94305-4085}
  \author{T.L. Duvall, Jr.}
  \institute{Laboratory for Astronomy and Solar Physics\\
NASA Goddard Space Flight Center, Greenbelt, MD 20771}
\end{opening}

\begin{document}
\begin{abstract}
We  present a new method for helioseismic diagnostics of the
three-dimensional structure of sound speed, magnetic fields and flow
velocities in the convection zone by inversion of acoustic travel-time
data.  The data are measurements of the time for acoustic waves to travel
between points on the solar surface and surrounding annuli obtained
from continuous observations at the South Pole in 1991 and from
high-resolution observations from the Solar and Heliospheric Observatory
(SOHO) in 1996.  The travel time of the waves depends primarily on the
sound speed perturbations and the velocity of flow along the ray paths.
The effects of the sound speed perturbations and flows can be separated
by measuring the travel time of waves propagating in opposite directions
along the same ray paths. Magnetic fields result in anisotropy of the wave
speed. A 3D inversion method based on Fermat's Principle and a regularized
least-squares technique have been applied to infer the properties of
convection in the quiet Sun and in active regions.

\end{abstract}
\vspace*{-19cm}\hspace*{-0.56cm}\tiny{SCORe'96 : Solar Convection and Oscillations and their Relationship, Proceedings of a workshop, held in Aarhus, Denmark, May 27 - 31, 1996, Eds.: F.P. Pijpers, J. Christensen-Dalsgaard, and C.S. Rosenthal, Kluwer Academic Publishers (Astrophysics and Space Science Library Vol. 225), p. 241-260, 1997} 
\vspace*{19cm}
\normalsize

\section{Introduction}
Observation of solar acoustic waves provides
important information about the internal structure and dynamics of
the Sun. The solar waves are observed as 5-min oscillations of Doppler
velocity or spectral-line intensity at the surface. The observations
are usually represented in terms of normal modes of the Sun. Most
research in helioseismology is focused on
determining the internal stratification and rotation by inverting
frequencies of the modes (Gough \& Toomre, 1991).
This traditional  approach has provided
interesting information about the radial structure, large-scale
asphericity and differential rotation. However, using mode frequencies
alone one could possibly study only azimuthally averaged, symmetrical
relative to the equator components of the structure and rotation, whereas
direct observations of the solar surface show substantial deviations from
this symmetry.
Therefore, considerable efforts in helioseismology are devoted to
local helioseismic diagnostics, such as the ring-diagram analysis
based on measuring the local dispersion relation of the acoustic
waves (Gough \& Toomre, 1983; Hill, 1988; Haber, {\it et al}, 1994;
Bogdan \& Braun, 1995), and the time-distance seismology
based on travel times of groups of modes  (Duvall, {\it et al.}, 1993, 1996;
D'Silva \& Duvall, 1995; Kosovichev, 1996).

In Section 2, we discuss some principles of  time-distance helioseismology,
interpretation of the travel time measurements in the acoustical ray
approximation and the role of wave effects. Our numerical inversion procedure
and test results are described in Section 3. In Section 4, we present
some first inferences about convection in the quiet Sun and in active regions.

\section{Time-distance seismology}
\subsection{Observation of the cross-covariance function}

The basic idea of time-distance helioseismology is to measure the
acoustic travel time between different points on the solar surface,
and then to use the measurement for inferring variations of the structure
and flow velocities in the interior along the wave paths connecting the
surface points. This idea is similar to the Earth's seismology.
However, unlike in Earth, the solar waves are generated stochastically
by numerous acoustic sources in the subsurface layer of turbulent
convection. Therefore, to measure the wave travel time,
we use the cross-covariance
function, $\Psi(\tau, \Delta)$, of the oscillation signal, $f(t,\vecr)$,
between different points on the solar surface (Duvall, {\it et al.}, 1993):
\begin{equation}
\Psi(\tau,\Delta) = \int_0^T f(t,\vecr_1) f^*(t+\tau,\vecr_2) dt,
\end{equation}
where $\Delta$ is the angular distance between the points with
coordinates $\vecr_1$ and $\vecr_2$, $\tau$ is the delay time, and
$T$ is the total time of the observations. Because of the stochastic
nature of excitation
of the oscillations, function $\Psi$ must be averaged over some
areas on the solar surface to achieve a signal-to-noise ratio
sufficient for measuring travel times $\tau$. The oscillation signal,
$f(t,\vecr)$, is usually the Doppler velocity or intensity. A typical
cross-covariance function averaged over the whole disk is shown in
 Figure 1. The observational
data were obtained at the South Pole on January 5, 1991, by Jefferies
{\it et al.} (1994). The oscillation signal was intensity variations
of Ca$^+$K line, filtered to select the p-mode frequency range
using Gaussian transfer function
\begin{equation}
G(\omega)=\exp\left[-\left(\frac{\omega-\omega_0}{\delta\omega}\right)^2\right],
\end{equation}
where $\omega$ is the cyclic frequency, $\omega_0$ is the central frequency
and $\delta\omega$ is the characteristic bandwidth of the filter.
The cross-covariance function in Fig. 1 displays three sets of ridges
which correspond to the first, second
and third bounces of packets of acoustic wave packets from the surface.

\begin{figure}[t]
\centerline{\includegraphics*[width=0.8\linewidth]{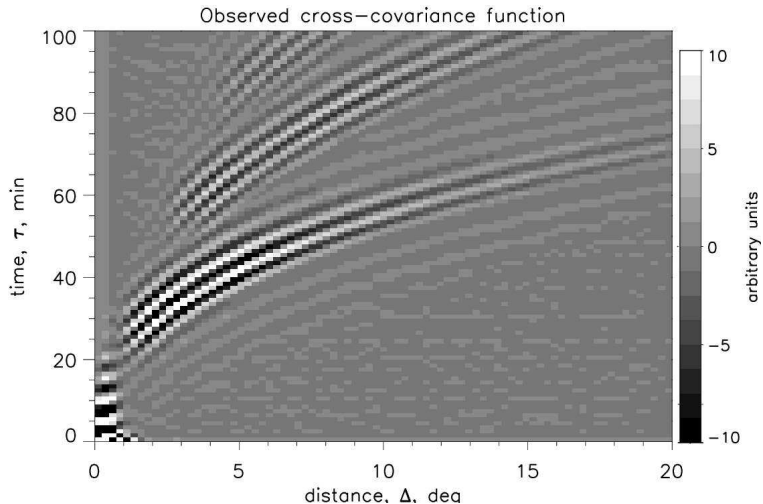}}
	\caption{
		The observed cross-covariance function inferred from the data obtained at the
		South Pole by Jefferies {\it et al.} (1994) as a function of distance on
		the solar surface, $\Delta$, and the delay time, $\tau$.}
\end{figure}

The cross-covariance function represents a solar `seismogram'. Ideally,
the seismogram should be inverted to infer the structure and flows
using a wave theory. However, in practice, like in terrestrial seismology
(e.g. Aki \& Richards, 1980) and oceanography (Munk, {\it et al.}, 1995),
we have to use a different approximation, the most simple and powerful of
which is currently the geometrical acoustic (ray) approximation.
In the next Section, we discuss relations between the modal wave approach
and the ray theory.

\subsection{Cross-covariance function and travel times}

Generally, the observed  solar oscillation signal corresponds
to radial displacement or pressure perturbation, and can be
represented in terms of normal modes, or standing waves
(e.g. Unno {\it et al.}, 1989):
\begin{equation}
f(t,r,\theta,\phi) = \sum_{nlm} a_{nlm}\xi_{nlm}(r,\theta,\phi)
\exp(i\omega_{nlm}t + i\phi_{nlm}),
\end{equation}
where $n, l$ and $m$ are the radial order, angular degree and
angular order of a normal mode respectively, $\xi_{nlm}(r,\theta,\phi)$
is a mode eigenfunction in the spherical coordinates, $r,\theta$ and $\phi$,
$\omega_{nlm}$ is the eigenfrequency, and $\phi_{nlm}$ is an initial
phase of the mode. Using equation (2), we express
the cross-covariance function in terms of normal modes, and then
represent it as a superposition of traveling wave packets.
Such a representation is important for interpretation of the time-distance
data, and for studying from the data the regional structures in the Sun.
The correspondence between the normal modes and the wave packets has been
discussed for surface oscillations in Earth's seismology
(e.g. Ben-Menahem, 1964) and also for ocean waves (e.g. Tindle \& Guthrie, 1974).

To simplify the analysis, we consider the spherically symmetrical case.
For a radially stratified sphere, the eigenfunctions can be represented
interms of spherical harmonics $Y_{lm}(\theta,\phi)$:
\begin{equation}
\xi_{nlm}(r,\theta,\phi) =  \xi_{nl}(r)Y_{lm}(\theta,\phi),
\end{equation}
where $\xi_{nl}(r)$ is the radial eigenfunction (e.g. Unno, {\it et al.},
1989).

Then, from the convolution theorem (e.g. Bracewell, 1986):
\begin{equation}
\Psi(\tau,\Delta) = \int_{-\infty}^\infty F(\omega,\vecr_1) F^*(\omega,\vecr_2)
\exp(i\omega\tau)d\omega,
\end{equation}
where $F(\omega,\vecr)$ is Fourier transform of the oscillation signal
$f(t,\vecr)$, filtered with the Gaussian transfer function (2).
The time series used in our analysis are considerably longer than the
travel time $\tau$, therefore, we can neglect the effect of the window function,
and represent $F(\omega,\vecr)$ in the form
\begin{equation}
F(\omega,r,\theta,\phi) \approx \sum_{nlm} a_{nl}\xi_{nl}(r)Y_{lm}(\theta,\phi)
\delta(\omega-\omega_{nl})\exp\left[-\left(\frac{\omega-\omega_0}{\delta\omega}\right)^2\right],
\end{equation}
where $\delta(x)$ is the delta-function, and $\omega_{nl}$ are frequencies
of the normal modes.
In addition, we assume the normalization conditions:  $\xi_{nl}(R)=1$,
$a_{nl}=1$.
Then, the cross-covariance function is
\begin{equation}
\Psi(\tau,\vecr_1,\vecr_2)=\sum_{nl}\sigma_{nl}\exp\left[-\left(\frac{\omega-\omega_0}
{\delta\omega}\right)^2 + i\omega_{nl}\tau\right]\sum_{m=-l}^{l}
Y_{lm}(\theta_1,\phi_1)Y^*_{lm}(\theta_2,\phi_2).
\end{equation}
The sum of the spherical function products
\begin{equation}
\sum_{m=-l}^{l}
Y_{lm}(\theta_1,\phi_1)Y^*_{lm}(\theta_2,\phi_2)=\alpha_lP_l(\cos\Delta),
\end{equation}
where $P_l(\cos\Delta)$ is the Legendre polynomial, $\Delta$ is the angular
distance between points 1 and 2 along the great circle on the sphere,
$\cos\Delta=\cos\theta_1\cos\theta_2+\sin\theta_1\sin\theta_2\cos(\phi_2-\phi_1)$,
and $\alpha_l=\sqrt{4\pi/(2l+1)}$.
Then, the cross-covariance function in terms of the normal modes
\begin{equation}
\Psi(\tau,\Delta) \approx \sum_{nl}a_{nl}\alpha_lP_l(\cos\Delta)
\exp\left[-\left(\frac{\omega_{nl}-\omega_0}
{\delta\omega}\right)^2+i\omega_{nl}\tau\right].
\end{equation}
\begin{figure}
\centerline{\includegraphics*[width=0.8\linewidth]{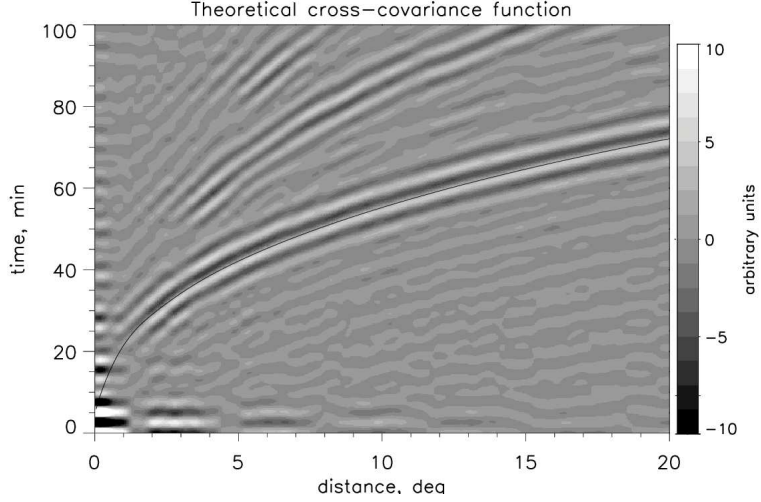}}
  \caption{
   The theoretical cross-covariance function of the solar p-modes of $l=0-1000$
as a function of distance on the solar surface, $\Delta$, and time, $\tau$.}
\end{figure}
We use Eq.(9) to generate the cross-covariance function for solar models.
An example of the theoretical cross-covariance function of p modes of the standard
solar model of Christensen-Dalsgaard, {\it et al.} (1996) is shown
in Fig. 2.

By grouping
the modes in Eq. (9) in narrow ranges of the angular phase velocity,
$v=\omega_{nl}/L$, where $L=\sqrt{l(l+1)}$,
 and applying the method of stationary phase,
the cross-covariance function can be approximately represented in the
form
\begin{equation}
\Psi(\tau,\Delta) \propto \sum_{\delta v}
\cos\left[\omega_0\left(\tau-
\frac{\Delta}{v}\right)\right]\exp\left[-\frac{\delta\omega^2}{4}\left(\tau-
\frac{\Delta}{u}\right)^2\right],
\end{equation}
where $\delta v$ is a narrow interval of the phase speed,
$u\equiv (\partial\omega/\partial k_h)$ is the horizontal component of the
group velocity, $k_h = 1/L$ is the angular component of the wave vector, and
$\omega_0$ is the central frequency of the frequency filter (2),
and $\delta\omega$ is the characteristic width of the filter.
We measure the  phase and group travel times by fitting
individual terms of Eq. (10), represented by a Gabor-type wavelet,
to the observed  cross-covariance
function using a least-squares technique. An example of the fit is shown
in Fig.3.
\begin{figure}
\centerline{\includegraphics*[width=0.8\linewidth]{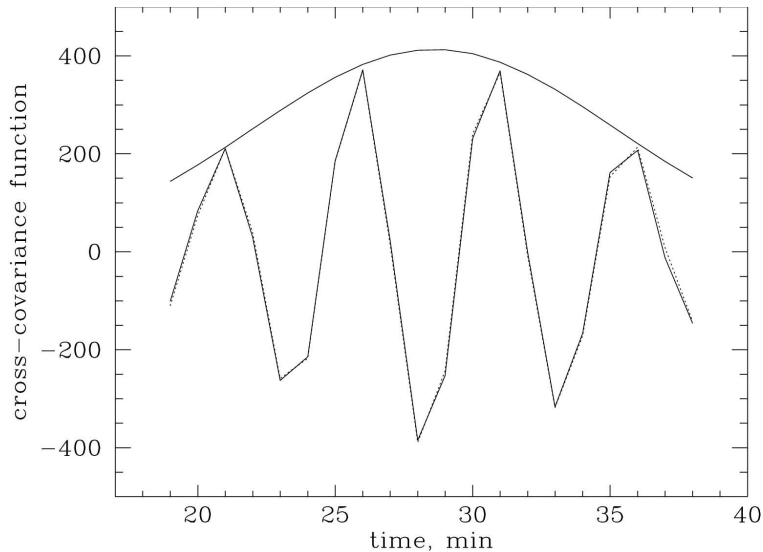}}
  \caption{
    An example of the Gabor wavelet  (Eq. 10)
fit to a cross-covariance function obtained
from the MDI data. The solid curves show the periodic component and the
envelope of the wavelet; the dashed curved is the measured cross-covariance
function.}
\end{figure}

This technique measures both phase ($\Delta/v$) and group
($\Delta/u$) travel times of the p-mode
wave packets. The previous time-distance measurements provided either
the group time (Jefferies, {\it et al.}, 1994), or an unspecified combination of
the group and phase times (Duvall,  {\it et al.}, 1996). We have found that
the noise level in the phase-time measurements is substantially
lower than in the group-time
measurements. Therefore, in this paper, we use the phase times.
We also use the geometrical acoustic (ray) approximation to relate the measured
phase times to the internal properties of the Sun. More precisely, we measure
the variations of the local travel times at different points on the surface,
relative to the travel times averaged over the observed area, and then
infer variations of the internal structure and flow velocities from the
travel time anomalies using a perturbation theory.

\subsection{Travel time perturbation}

In the ray approximation, the travel times are sensitive only to the
perturbations along the ray paths  given by
\begin{equation}
\frac{d\vecr}{dt} = \frac{\partial\omega}{\partial\veck}, \;\;\;\;\;
\frac{d\veck}{dt} = \frac{\partial\omega}{\partial\vecr},
\end{equation}
where $\vecr$ is the radius-vector and $\veck$ is the wave-vector.
The variations of the travel time obey
Fermat's Principle (e.g. Gough, 1993)
\begin{equation}
\delta\tau = \frac{1}{\omega}\int_\Gamma \delta\veck d\vecr,
\end{equation}
where $\delta\veck$ is the perturbation of the wave vector due to the
structural inhomogeneities and flows along the unperturbed ray path, $\Gamma$.

The dispersion relation for acoustic waves in the convection zone is
\begin{equation}
(\omega - \veck\vecU)^2 = \omega_c^2 + \veck^2 c_f^2,
\end{equation}
where $\vecU\;\;$ is the flow velocity, $\omega_c\;\;$ is the acoustic cut-off
frequency, $\;\;c_f^2=\frac{1}{2}
\left(c^2+c_A^2+\sqrt{\left(c^2+c_A^2\right)^2-
4c^2(\veck\veccA)^2/k^2}\right)$ is the fast magnetoacoustic speed,
$\veccA = \vecB/\sqrt{4\pi\rho}\;\;$ is the vector Alfv\'en velocity,
$\vecB$ is the magnetic field strength,
$c$ is the adiabatic sound speed, and $\rho$ is the plasma density.
If we assume that, in the unperturbed state $\vecU=\vecB=0$, then,
to the first-order approximation
\begin{equation}
\delta\tau=-\int_\Gamma\left[\frac{(\vecn\vecU)}{c^2} + \frac{\delta c}{c}S
+ \left(\frac{\delta\omega_c}{\omega_c}\right)
\frac{\omega_c^2}{\omega^2c^2S} +\frac{1}{2}\left(\frac{c_A^2}{c^2}
-\frac{(\veck\veccA)^2}{k^2c^2}\right)S\right]ds,
\end{equation}
where $\vecn$ is a unit vector tangent to the ray, $S=k/\omega$ is the
phase slowness. Then, we separate the effects of flows and structural
perturbations by taking the difference and the mean of the reciprocal travel
times:
\begin{equation}
\delta\tau_{\rm diff} = - 2\int_\Gamma\frac{(\vecn\vecU)}{c^2}ds;\\
\end{equation}
\begin{equation}
\delta\tau_{\rm mean} =-\int_\Gamma\left[\frac{\delta c}{c}S +
\left(\frac{\delta\omega_c}{\omega_c}\right)
\frac{\omega_c^2}{\omega^2c^2S} +\frac{1}{2}\left(\frac{c_A^2}{c^2}
-\frac{(\veck\veccA)^2}{k^2c^2}\right)S\right]ds.
\end{equation}
Anisotropy of the last term of Eq. (16) allows us to separate, at least
partly, the magnetic effects from the variations of the
sound speed and the acoustic cut-off frequency. The acoustic cut-off frequency,
$\omega_c$ may be perturbed by the surface magnetic fields and by
the temperature and density inhomogeneities. The effect of the
cut-off frequency variation depends strongly on the wave frequency,
and, therefore, it should result in frequency dependence in $\tau_{\rm mean}$.
However, we have not detected yet a significant frequency dependence in the
observed travel times.

Typically, we measure times for acoustic waves to travel between
points on the solar surface and surrounding quadrants symmetrical
relative to the North, South, East and West directions.
In each quadrant, the travel times are averaged over narrow ranges of
travel distance $\Delta$.
Then, the times for northward-directed waves are subtracted from
the times for south-directed waves to yield the time, $\tau_{\rm diff}^{\rm NS}$,
which predominantly measures north-south motions. Similarly, the time
differences, $\tau_{\rm diff}^{\rm EW}$, between westward- and eastward directed
waves yields a measure of east-ward motion. The time,  $\tau_{\rm diff}^{\rm oi}$,
between outward- and inward-directed waves, averaged over the full annuli,
is mainly sensitive to vertical motion and the horizontal divergence.
This approximate separation onto the flow components
provides a qualitative picture of the motions on
the scale larger than the size of the quadrants, and is useful for a
preliminary analysis. However, in our numerical inversion of the data (Sec. 3),
all three components of the flow velocity are properly taken into account.
The averaging procedure described here is essential for reducing noise in
the data.

In addition, a similar averaging procedure is applied for $\tau_{\rm mean}$
to measure the times corresponding to the isotropic and anisotropic
components of the
structural perturbations. While the anisotropy can be estimated from
the quadrant-averaged data, to separate the magnetic field components,
$\tau_{\rm mean}$ is averaged over octants. We have found that the
anisotropic component in the quiet Sun is substantially weaker than
the isotropic component, even around magnetic structures of the chromospheric
network.

\subsection{Wave effects}
Obviously, we are concerned about the accuracy of the ray approximation,
particularly, for short travel distances and small-scale perturbations.
Because a wave theory for the time-distance seismology of the 3-D convective
structures has not been developed, the precise limits of the spatial
resolution of the ray approximation are not determined.
 However, we have studied the the resolution with depth, which is one
of the main concerns, in particular, for short travel distances
when the wavelength becomes comparable with the characteristic scale.
In this study, we have carried out a 'hare-and-hounds' test
 for spherically symmetrical perturbations of the sound speed.
 The test procedure was the following. One of us (AGK) computed
the theoretical cross-covariance functions, as described in Sec. 2.2,
for a standard solar model and for the same model with a localized sound-speed
perturbation,
\begin{equation}
\delta c/c = A\exp [-0.5(r-r_0)^2/w^2],
\end{equation}
where $A$ is the amplitude of the perturbation, $r_0$ is the central
radius, and $w$ is the characteristic width of the perturbation.

$A=0.01, r_0=0.98R$, and $w=0.005R$. Then, the travel time difference
between the cross-covariance functions of the two models were determined by
the second author (TLD) without knowing the
characteristics of the perturbation. The result of his measurements
shown in Fig. 4 (solid curve) is in agreement with the travel time variation
estimated from the ray theory. Evidently, the structure perturbations
inferred in the ray approximation from the cross-covariance function
will appear somewhat smoother than they are in reality.

\begin{figure}
\centerline{\includegraphics*[width=0.8\linewidth]{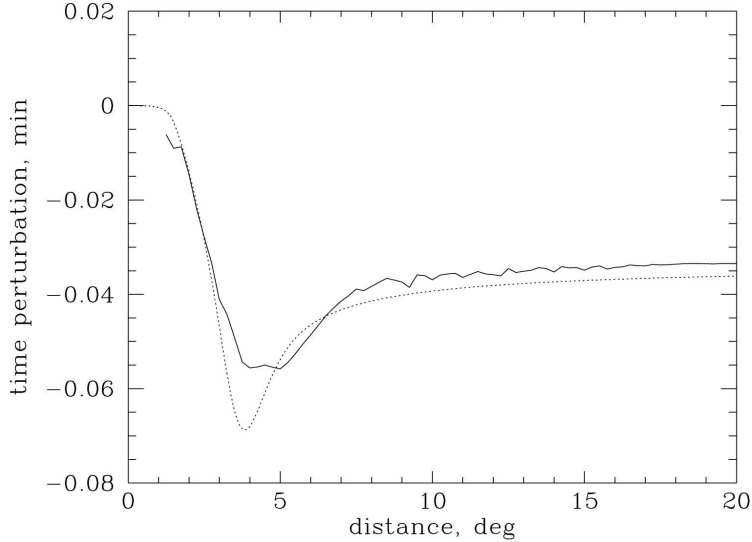}}
  \caption{
The travel time variation of the sound-speed perturbation (Eq. 17) with
$A=0.01, r_0=0.98R$, and $w=0.005R$, estimated
from the cross-covariance function (solid curve) and from the ray theory
(dotted curve).}
\end{figure}

Because of the wave effects, the travel time measured from the cross-covariance
function depends not only on the perturbations along a ray path but on the
perturbations in some vicinity of the ray path (cf Stark \& Nikolayev, 1993;
Bogdan, 1996).
This is illustrated
in Fig 5 which shows the energy density of a wave packet constructed using
a normal-mode solution to the solar oscillation equations, and corresponding
ray paths at three frequencies, 2.8, 3 and 3.5 mHz. The spreading along
the ray path is partly taken into account in our analysis by averaging
the travel times over a range of the travel distance. Beside the spreading,
the wave packet also shows some patchy interference structure with depth,
which is because the wave packet consists only of the modes of the radial
order $n$ of 1, 2 and 3.

\begin{figure}
	\centerline{\includegraphics*[width=0.8\linewidth]{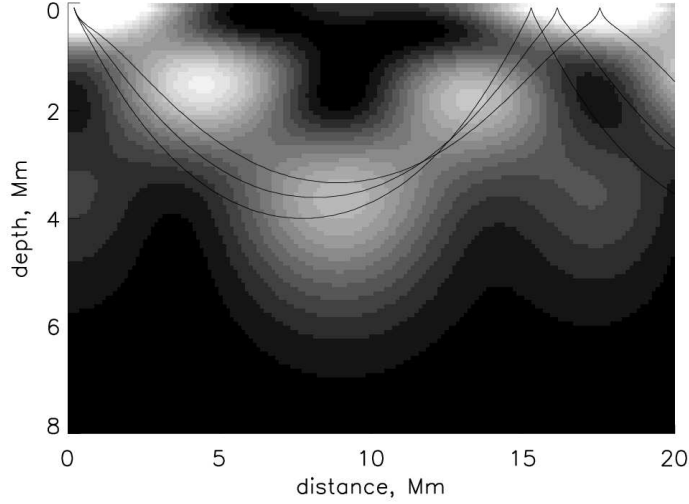}}
  \caption{
   The energy density as a function of the depth and distance of a wave
packet (gray-scale image), and the corresponding ray paths of the frequencies
2.8, 3, and 3.5 mHz.)}
\end{figure}
 \begin{figure}
 	\centerline{\includegraphics*[width=0.8\linewidth]{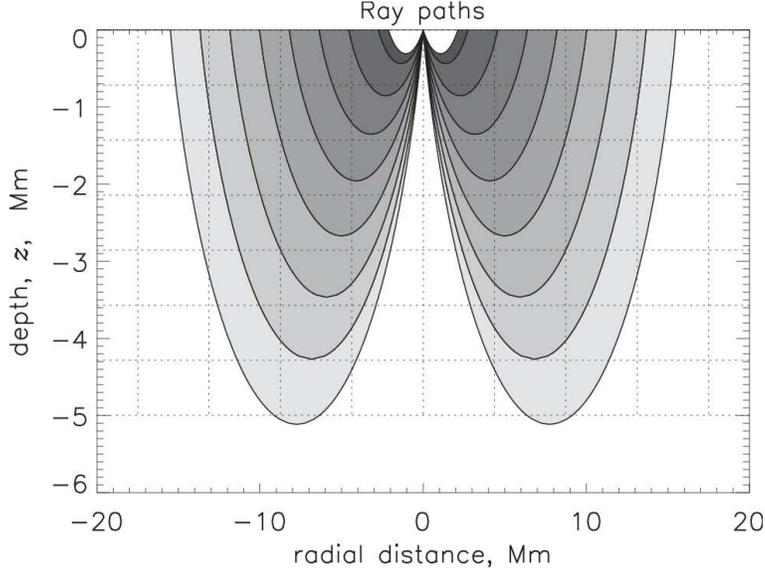}}
  \caption{
     The regions of ray propagation (shaded areas)
  as a function of depth, $z$,
and the radial distance, $\Delta$, from a point on the surface. The
rays are also averaged over a circular regions on the surface, forming
three-dimensional figures of revolution.
The dashed lines are the boundaries
of the cells of the inversion model.}
\end{figure}

\section{Inversion method}

\subsection{Discrete model}

We assume that the convective structures and flows do not change during the
observations and represent them by a discrete model. In this model,
the 3-D  region of wave propagation is divided into rectangular blocks.
The perturbations of the sound speed and the three components
of the flow velocity
are approximated by linear functions of coordinates within each block, e.g.
\begin{equation}
\vecU(x,y,z)=\sum\vecC_{ijk}\left[1-\frac{|x-x_i|}{x_{i+1}-x_i}\right]
\left[1-\frac{|y-y_j|}{y_{j+1}-y_j}\right]
\left[1-\frac{|z-z_k|}{z_{k+1}-z_k}\right],
\end{equation}
where
$x_i, y_j, z_k$ are the coordinates of the rectangular grid,
$\vecC_{ijk}$ are the values of the velocity in the grid points, and
$x_{i-1}\leq x\leq x_{i+1}$, $y_{j-1}\leq y\leq y_{j+1}$, and
$z_{k-1}\leq z\leq z_{k+1}$.
A part of the $x-z$ grid is shown in Fig.6
together with the ray system used in the inversions.

The travel time measured at a point on the solar surface is the result of the
cumulative effects of the perturbations in each of the  traversed rays
of the 3D ray systems. Figure 6 shows, in the ray approximation,
the sensitivity to given subsurface location for a certain point on the
surface.
This pattern is then moved around for different surface points
in the observed area, so that overall the data are sensitive to all
subsurface points in the depth range 0-5 Mm.

Therefore, we average the
equations over the ray systems corresponding to the different
radial distance intervals of the data, using approximately the same
number of ray paths as in the observational procedure. As a result, we
obtain two systems of linear equations that relate the data to the sound
speed variation and to the flow velocity, e.g. for the velocity
field,
\begin{equation}
\delta\tau_{{\rm diff};\lambda\mu\nu}=
\sum_{ijk}\vecA_{\lambda\mu\nu}^{ijk}\cdot\vecC_{ijk},
\end{equation}
where vector-matrix $\vecA$ maps the structure properties into the
observed travel
time variations, and indexes $\lambda$ and $\mu$ label
 the location of the central point of a ray
system on the surface, and index $\nu$ labels surrounding annuli. These equations
are solved by a regularized least-squares technique (Paige \& Saunders, 1982).
\begin{figure}
\centerline{\includegraphics*[width=0.8\linewidth]{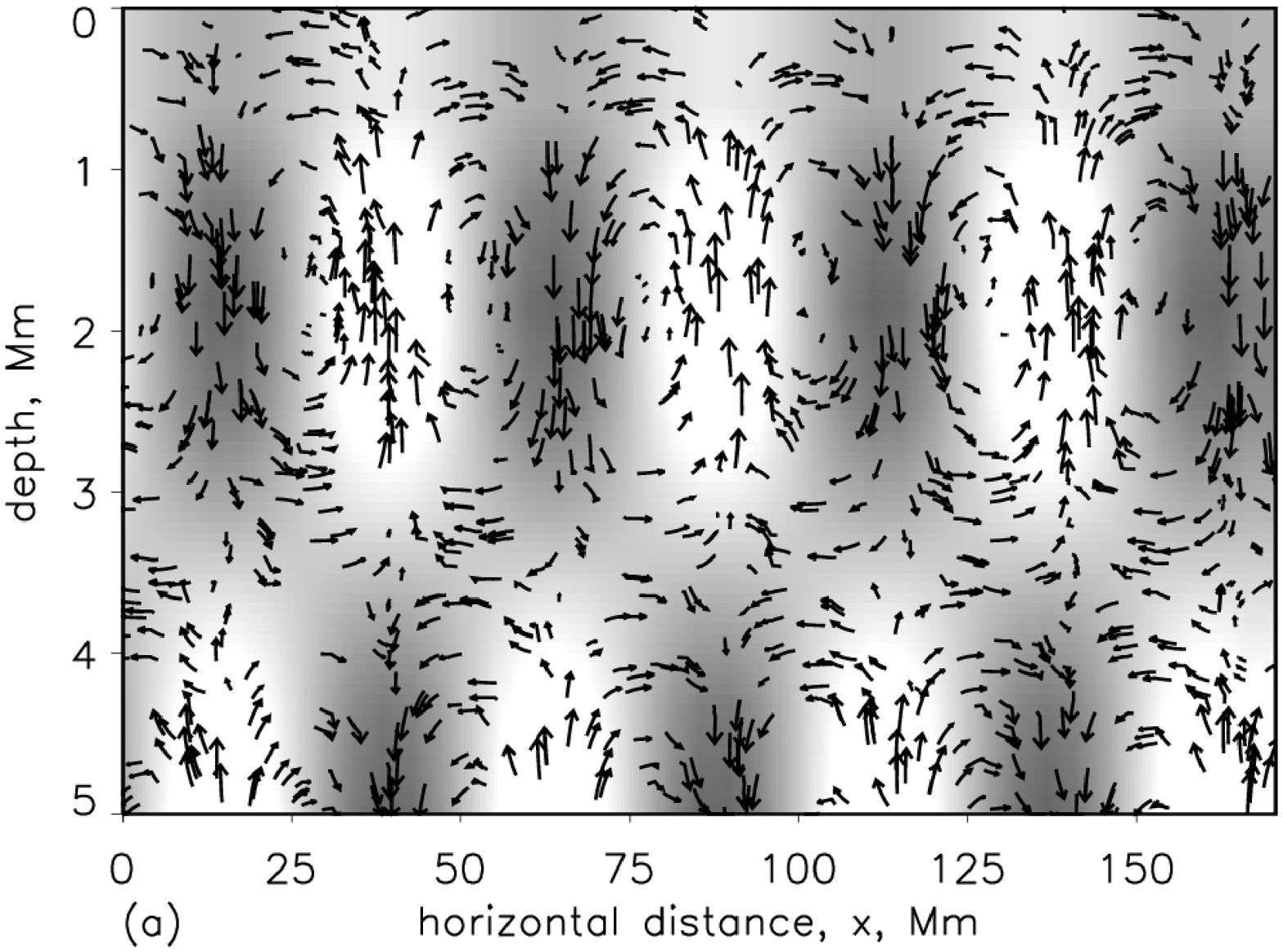}}
\centerline{\includegraphics*[width=0.8\linewidth]{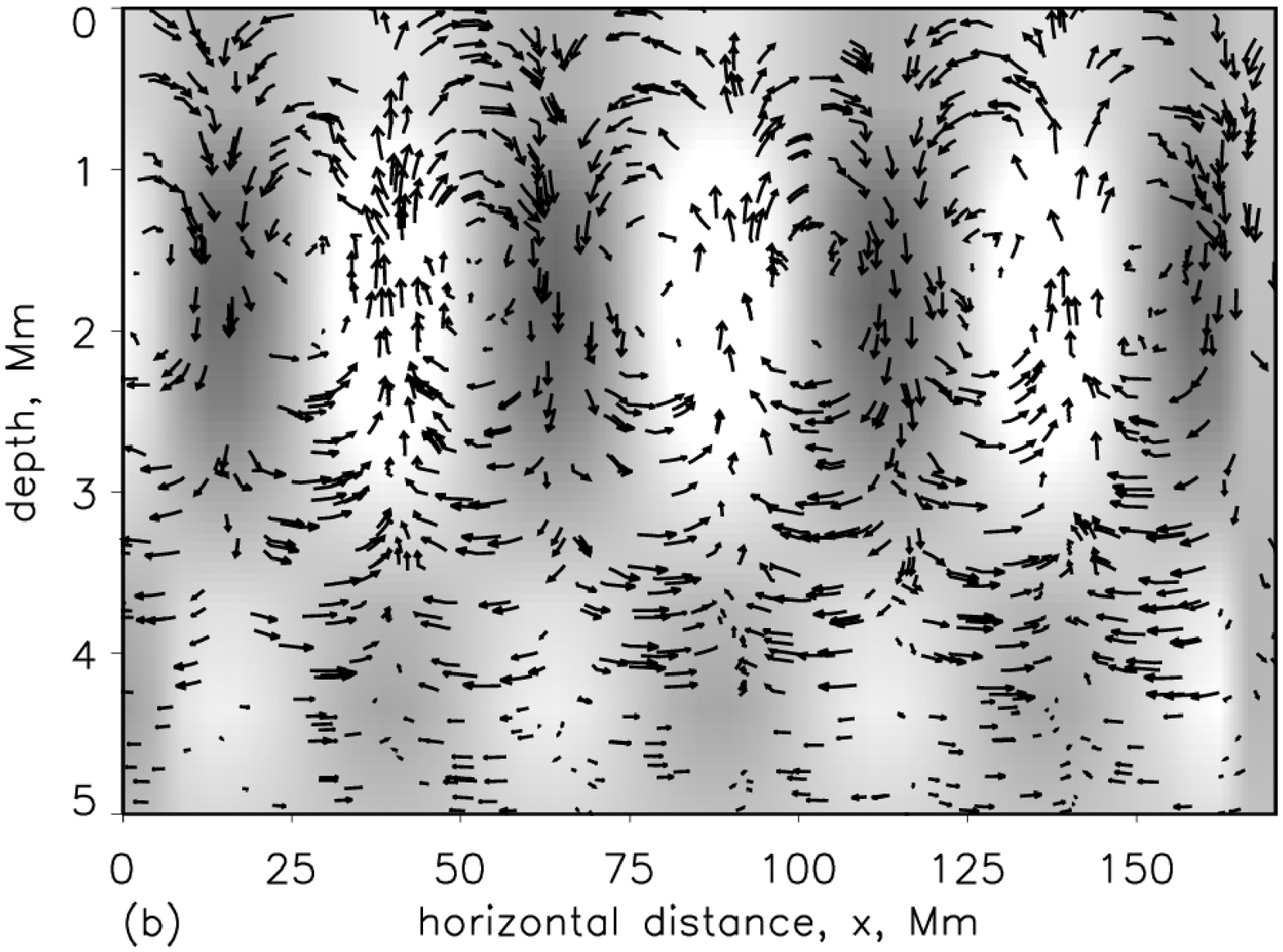}}
      \caption{
    (a) A vertical cut of the flow pattern (arrows) and
the sound-speed perturbation
(grey-scale background) of the test model of convection;
(b) the result of inversion of the travel times computed
for the system of rays shown in Fig.6.}
\end{figure}
\begin{figure}
	\centerline{\includegraphics*[width=0.8\linewidth]{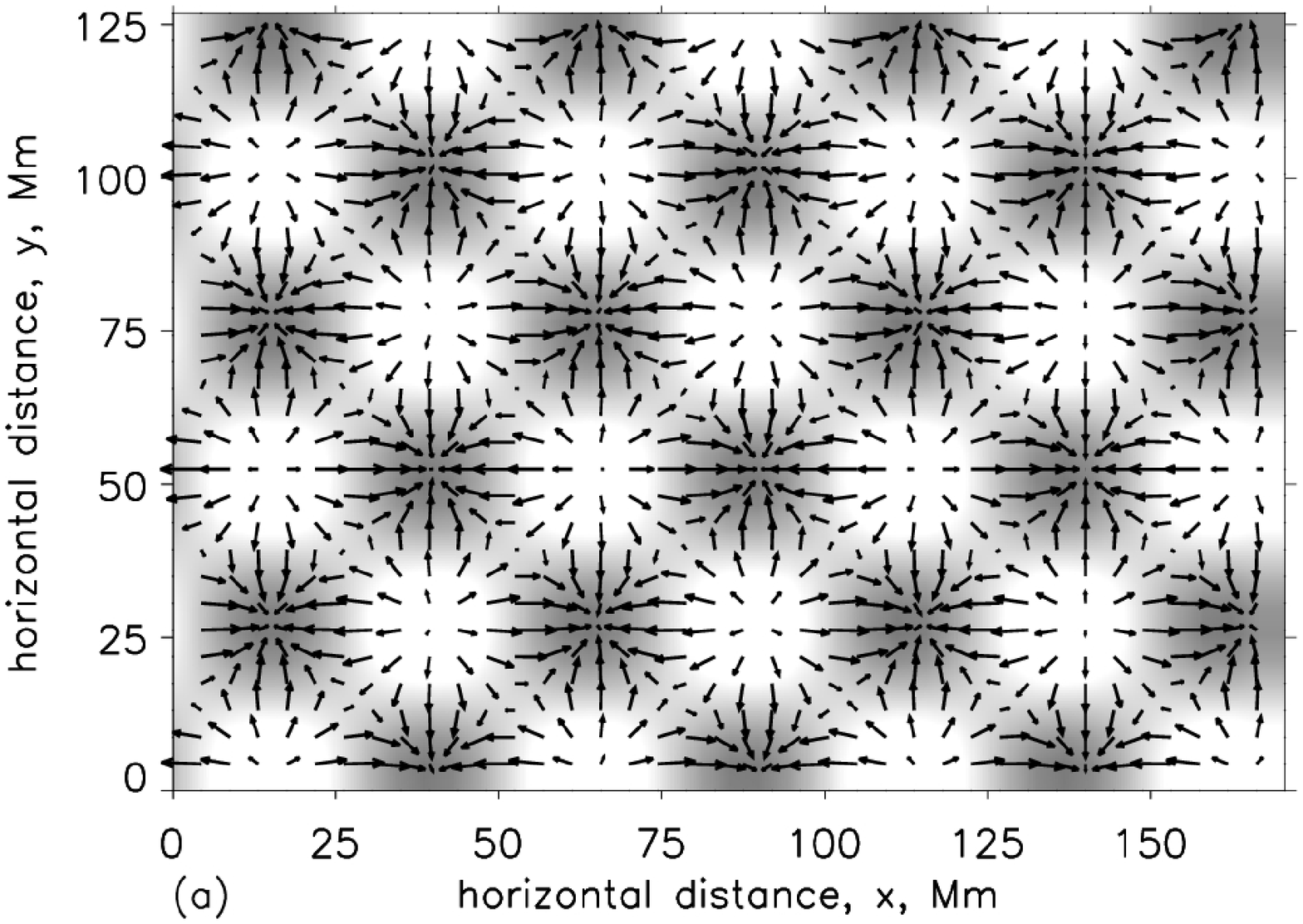}}
	\centerline{\includegraphics*[width=0.8\linewidth]{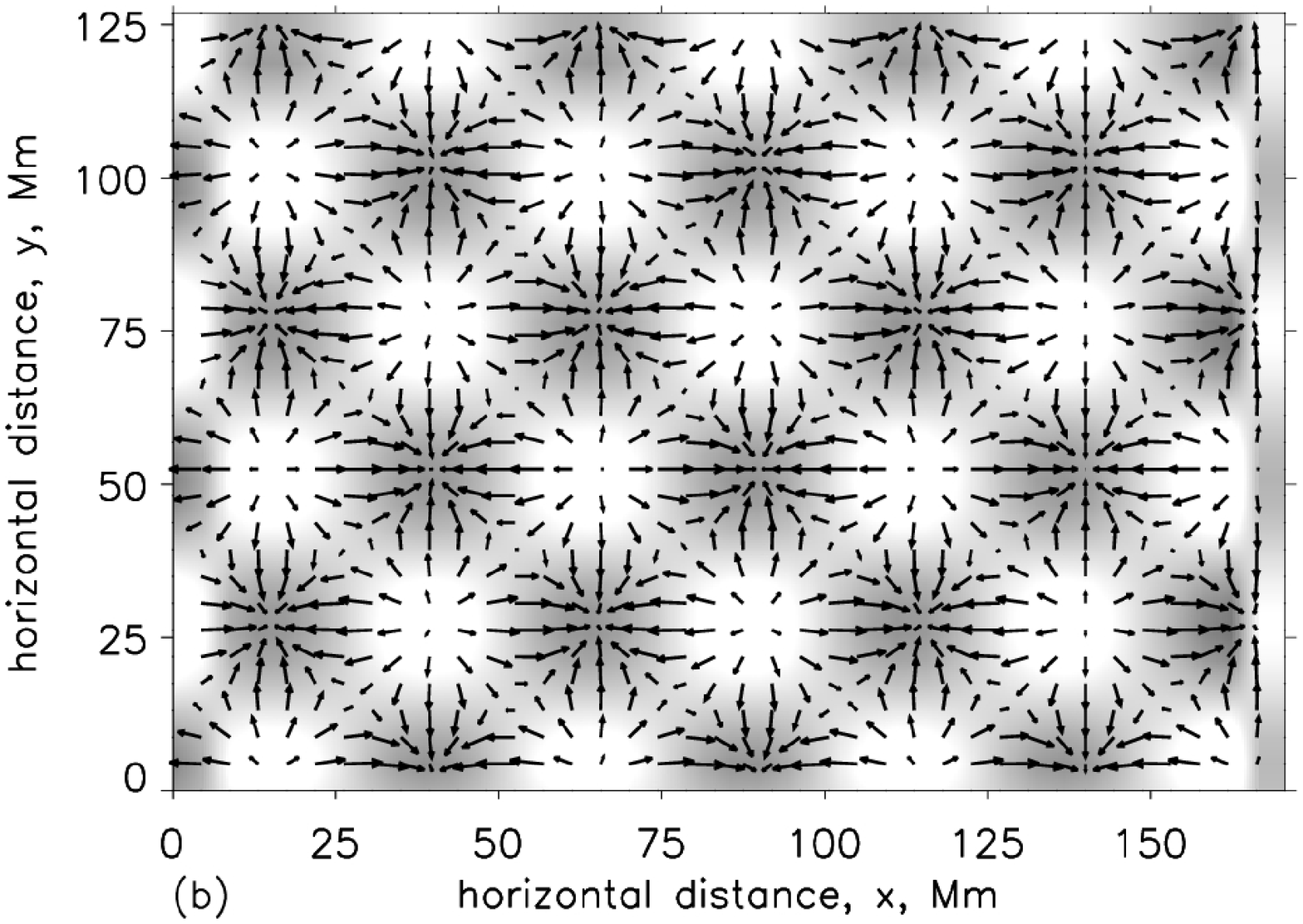}}
      \caption{
       (a) The horizontal flows (arrows) and
the sound-speed perturbation
(grey-scale background) of the test model of convection at depth 4.2 Mm;
(b) the result of inversion of the travel times computed
for the system of rays shown in Fig.6.}
\end{figure}

\subsection{Resolution and accuracy}
To test the resolution and accuracy of the inversions we took a simple
model of multi-level convective flow, and computed the averaged travel times
for the ray system shown in Fig.6 in the ray approximation.

The results of inversion of the test data in comparison with the original
model are shown in Figures 7 and 8.
These results demonstrate a very accurate reconstruction of the horizontal
components of the flow. However, the vertical flow in the deep layers
is not resolved because of the predominantly horizontal propagation
of the rays in these layers. The vertical velocities are also systematically
underestimated by 10-20\% in the upper layers.

\section{Initial inversion results}
\subsection{Quiet-Sun convection}
\begin{figure}
\centerline{\includegraphics*[width=0.95\linewidth]{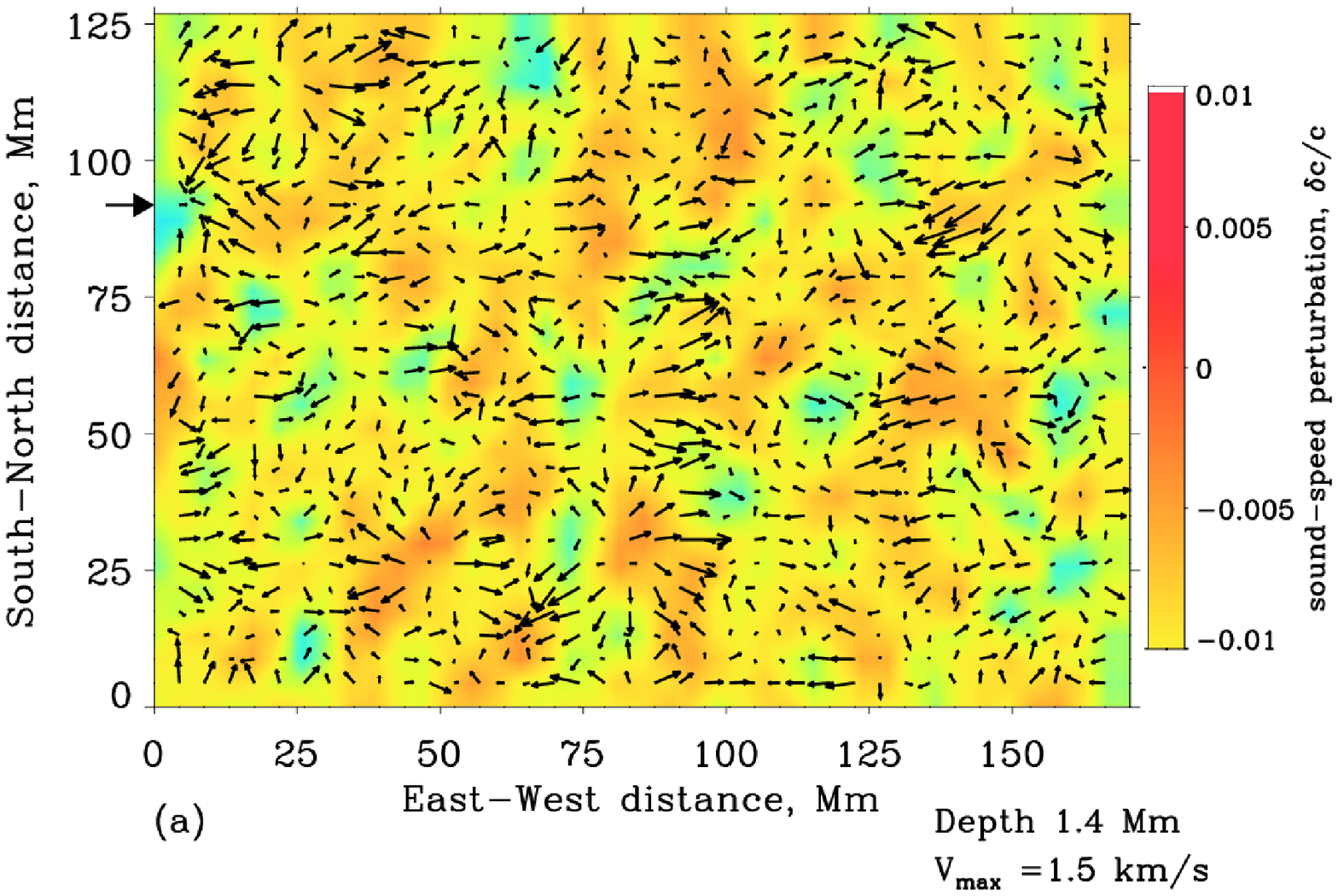}}
\centerline{\includegraphics*[width=0.95\linewidth]{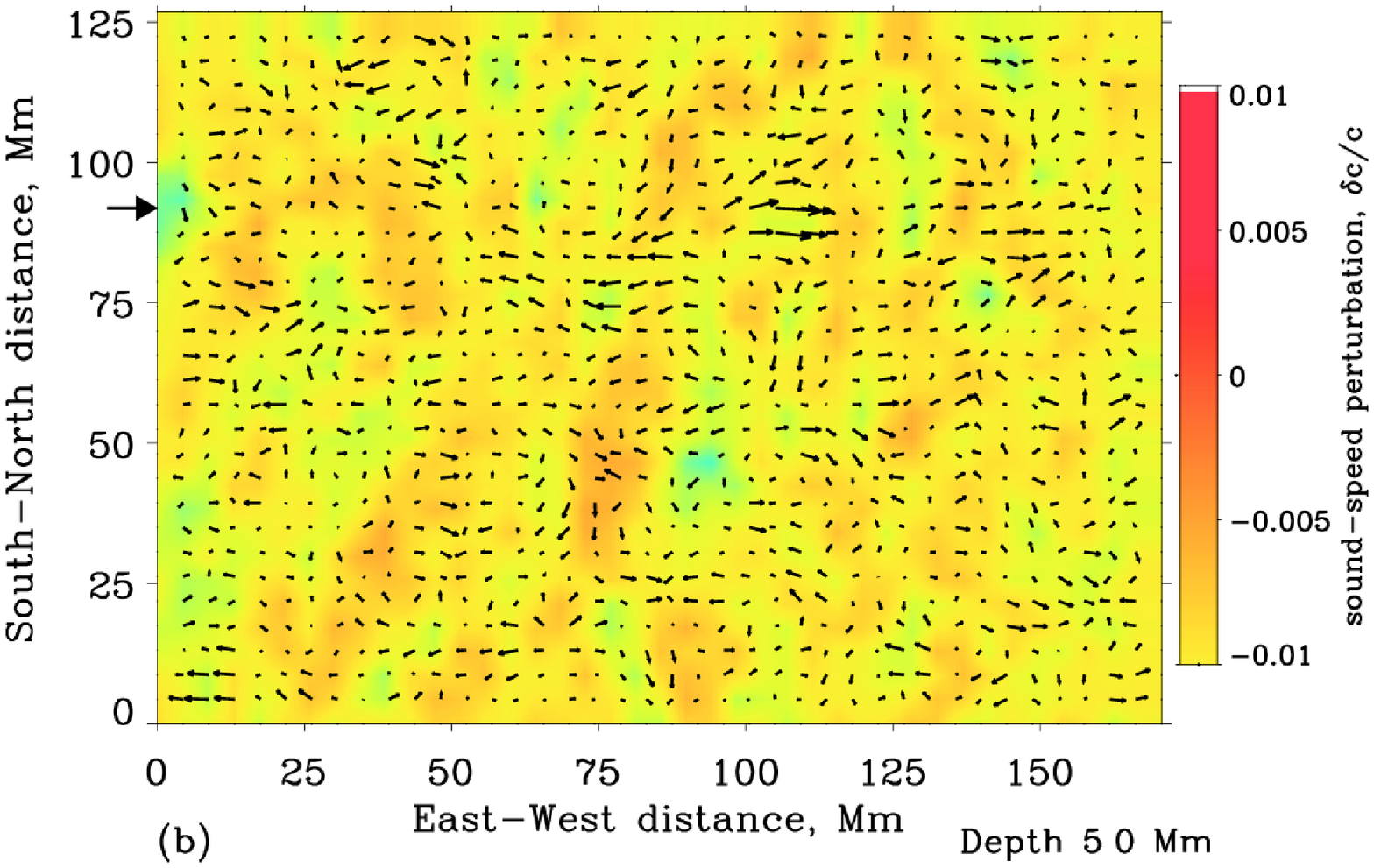}}
  \caption{
          The horizontal flow velocity field (arrows) and the sound-speed
perturbation  (grey-scale background) at the depths of 1.4 Mm (Fig. 9a) and
5.0 Mm (Fig. 9b), as inferred from the SOHO/MDI high-resolution data of
27 January 1996. The arrows at the South-North axis indicate
location of the vertical cut in East-West direction, which is shown
in Fig. 10.}
\end{figure}

\begin{figure}[t]
\centerline{\includegraphics*[width=0.95\linewidth]{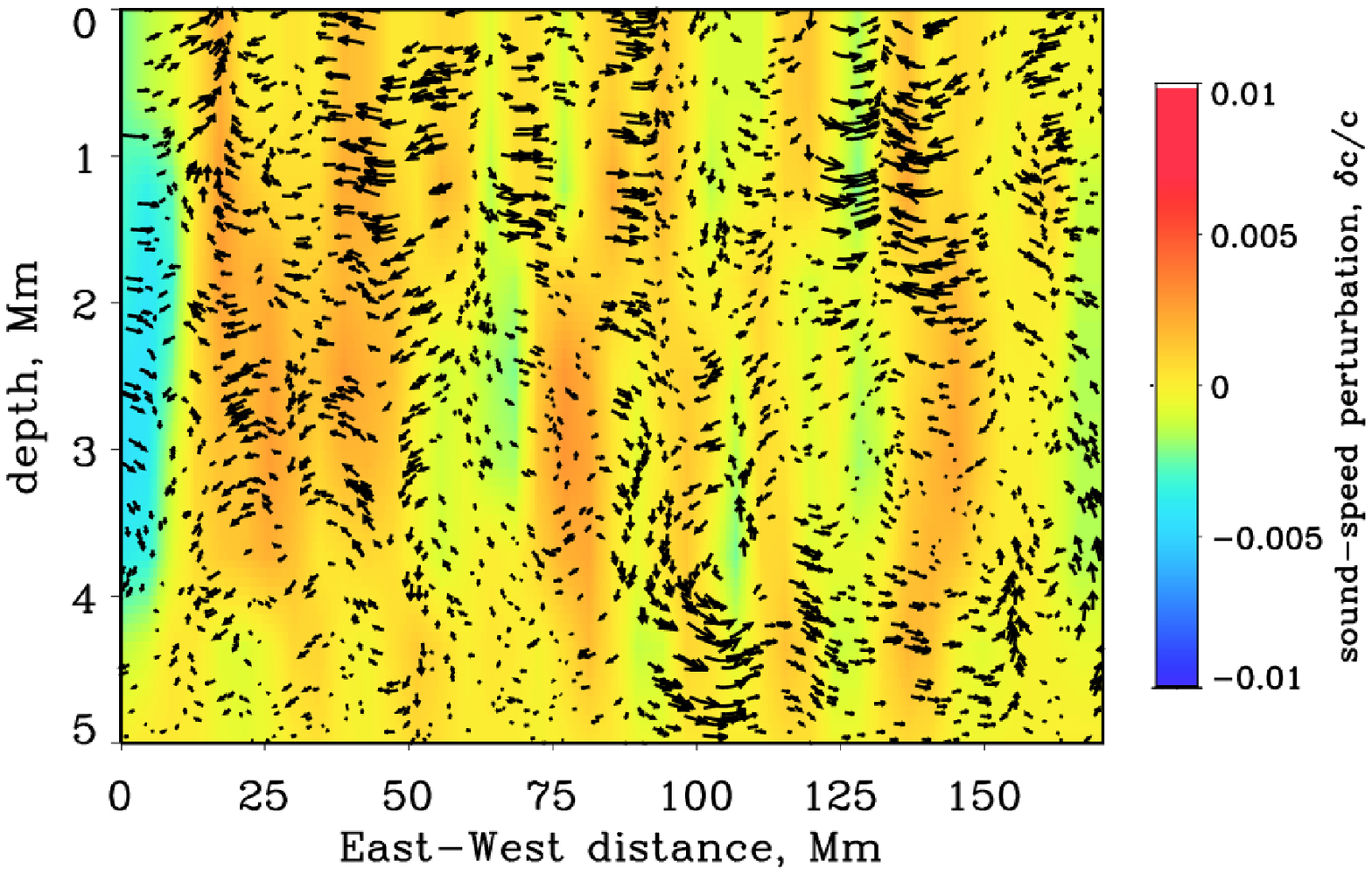}}
  \caption[]{
    The vertical flow field (arrows) and the sound-speed
perturbation  (grey-scale background) at the North-South position indicated
by arrows in Fig. 9.}
\end{figure}
Using the techniques of time-distance helioseismology, we have investigated
near-surface convective flows and structures using Doppler shift
data from the SOI/MDI experiment on SOHO.
The data used were for 8.5 hours on 27 January, 1996 from the
high resolution mode of the MDI instrument.
In this mode, the image sampling is 0.44 Mm per pixel,
and the image quality is excellent, as compared
with groundbased observations.   This enables us
to study the subsurface structure of convection
seen at the surface. Signal-to-noise ratio is enhanced
at the expense of spatial resolution
by averaging the cross-covariance function in spatial blocks of
$0.36^\circ\times 0.36^\circ$.
 The data are remapped onto a longitude,
 sine of latitude coordinate system.  A solid body rotation at the Carrington
rate is removed by shifting the pictures in longitude. To isolate p-mode waves,
the Gaussian filter (2) of full width at half-maximum 2 mHz and centered at
3.5mHz is applied to the temporal signal at each resultant point.

The travel time is determined by measuring displacements of the ridges of
the cross-covariance function at different points on the solar surface,
as described in Sec. 2. To examine waves at distances $\Delta$
shorter than 1 degree,
a spatial-frequency filter was applied
that multiplied the spatial Fourier transform by $k_h^{1.5}$, where $k_h$ is the
horizontal wave number.

The results of inversion of the data are shown in Figures 9 and 10.
We have found that, in the upper layers, 2-3 Mm deep, the horizontal flow
is organized in supergranular cells, with outflows from the center of
the supergranules. The characteristic size of the cells is 20-30 Mm.
Comparing with MDI magnetograms, we have found  that the cells
boundaries coincides with the areas of enhanced magnetic field.
These results are consistent with the observations of supergranulation
on the solar surface (e.g. Title, {\it et al.}, 1989).
 However, in the layers deeper than
2-3 Mm, we do not see the supergranulation pattern
which suggests that supergranulation is a superficial phenomenon.
The vertical flows (Fig. 10) correlate with the supergranular pattern
in the upper layers. Typically, there are upflows in the `hotter'
areas where the sound speed is higher than average, and downflows
in the `colder' areas.

These initial results look very promising for studying supergranulation
which certainly plays a very important role in the global dynamics of
the Sun, but have not been understood. Our results question the original
idea by Simon and Leighton (1964) that supergranulation represents large-scale
convective cells driven by convective instability in the HeII ionization
zone and, therefore, is, at least, 13 Mm deep. We hope to be able to observe
the flow pattern at this depth and investigate how it develops with time.

\begin{figure}[t]
\centerline{\includegraphics*[width=0.87\linewidth]{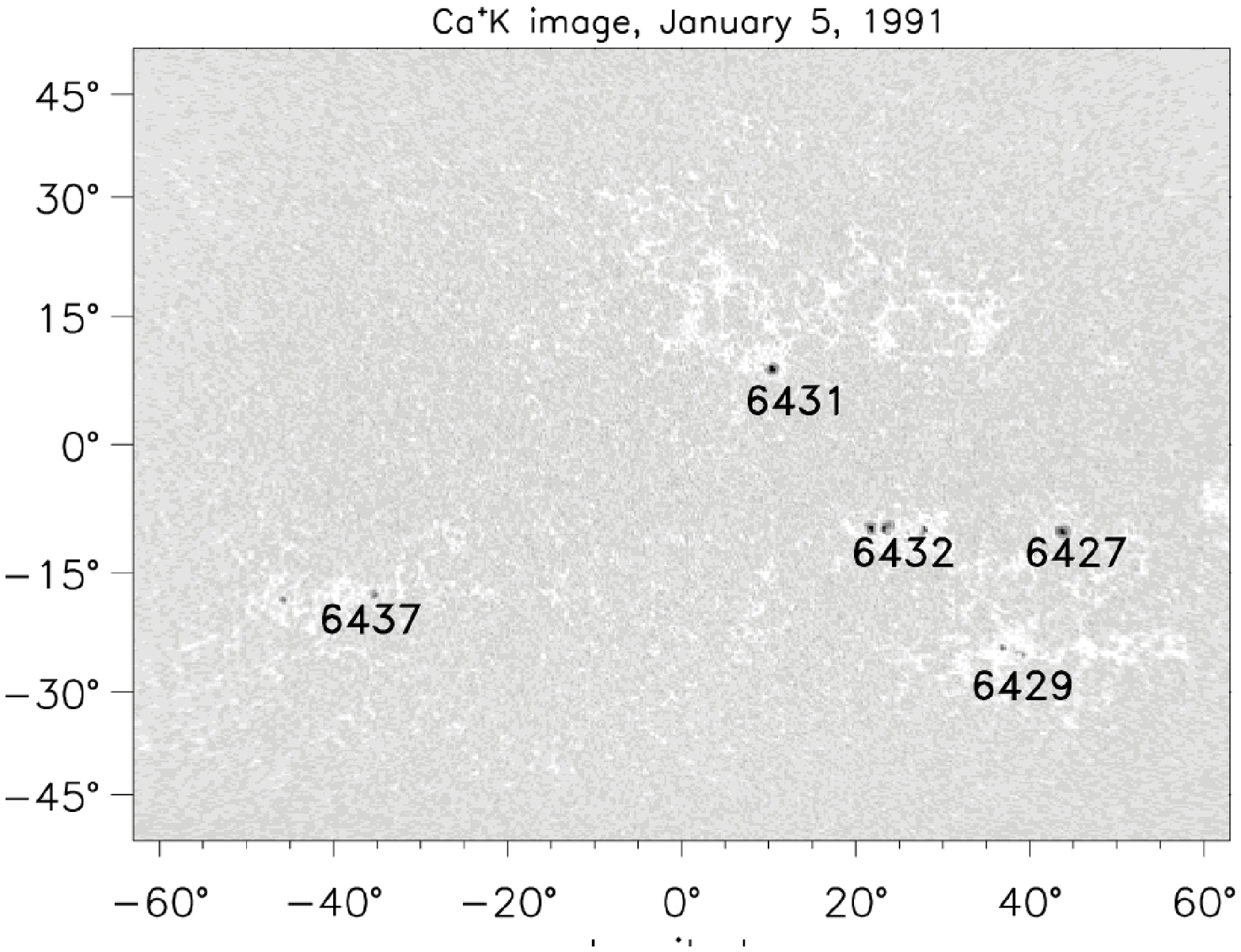}}
\centerline{\includegraphics*[width=0.95\linewidth]{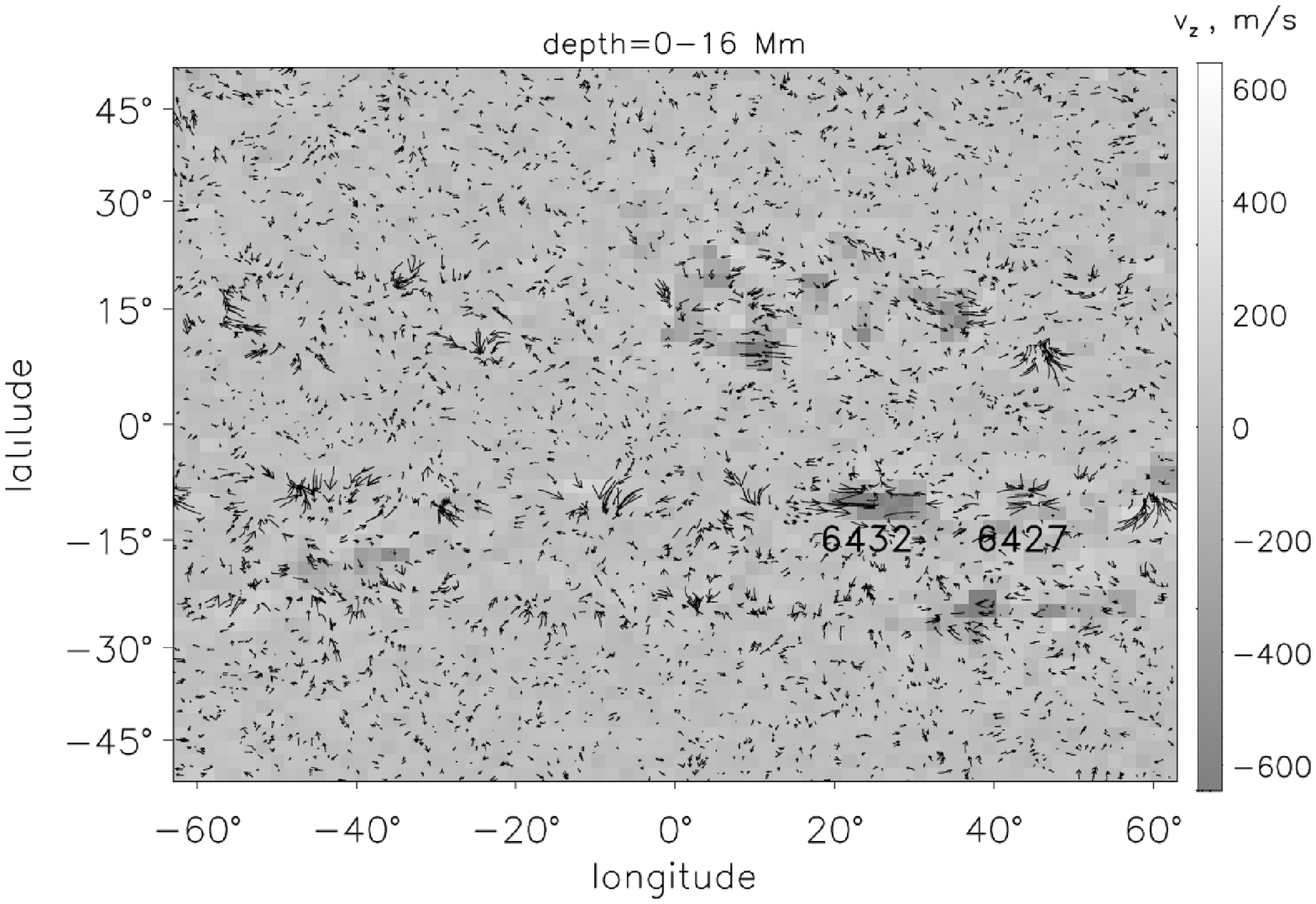}}
  \caption{
(a) Ca$^+$K-line image of the observed area observed by Jefferies {\it et al.}
(1994) on 5 January 1991 at the geographical South Pole;
(b) The horizontal (arrows) and vertical (grey-scale background) flow velocities
in the subsurface layer of the area shown in Fig. 11a. (after Kosovichev, 1996)}
\end{figure}
\subsection{Large-scale structures and flows in active regions}

\begin{figure}[t]
\centerline{\includegraphics*[width=0.95\linewidth]{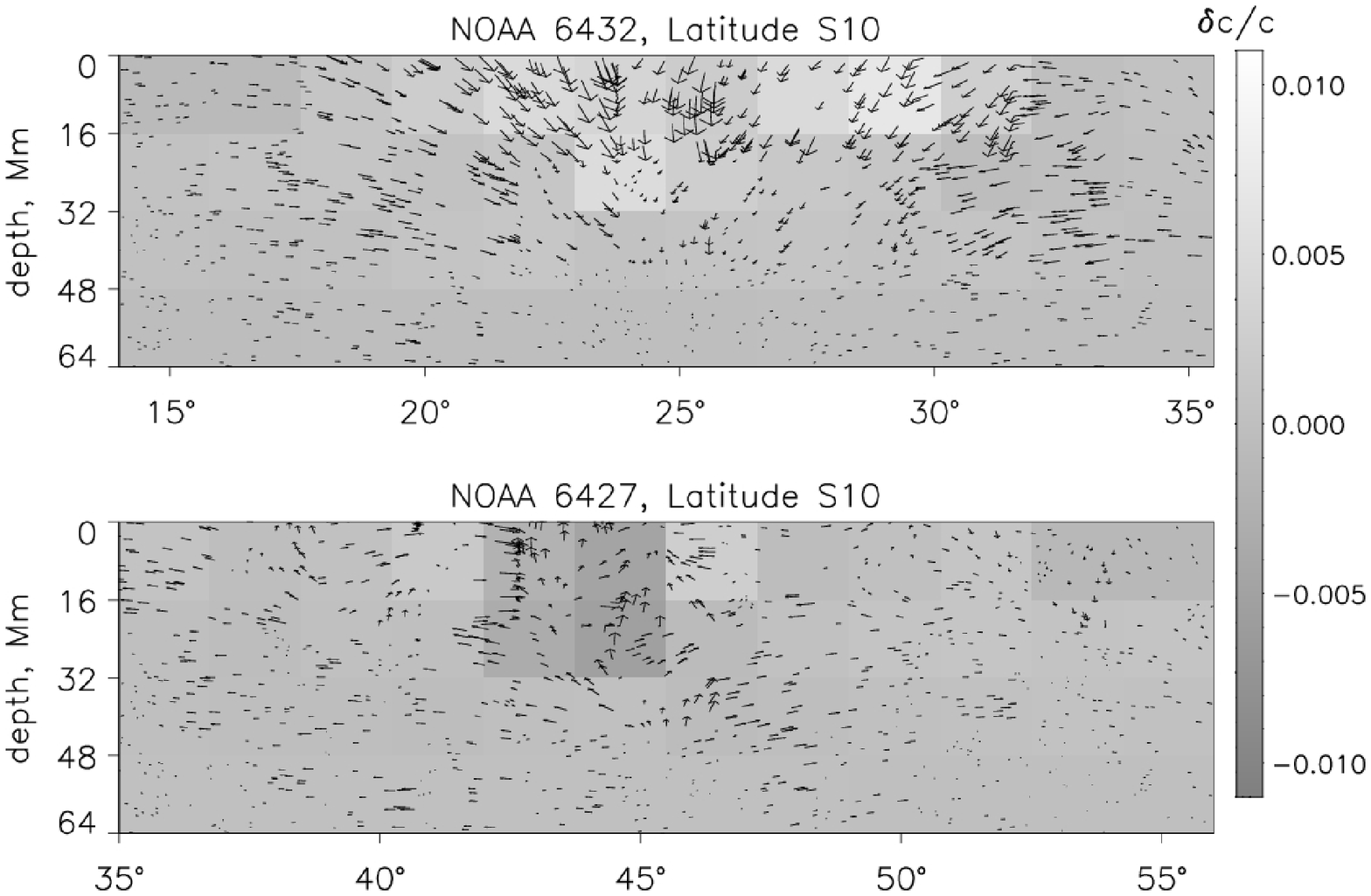}}
  \caption[]{
   Variations of the sound speed (grey-scale map) and the flow velocity
(arrows) as a function of depth and longitude at latitude $-10^\circ$
in growing active region NOAA 6432 (top), and in decaying region NOAA 6427
(bottom). ( Kosovichev, 1996)}
\end{figure}

There is no high-resolution data available yet to study convection of the
active Sun. However, we have studied some large-scale properties of
convection using the data obtained at the South Pole by Jefferies {\it
et al.} (1994) in January 1991, when the solar activity was high.
The observed area is shown in the Ca$^+$K-line intensity map in Fig. 11a.
The travel times in this area have been measured by Duvall {\it et al.} (1996)
for four sets of rays propagating from each of $72\times 52$ points
equally spaced in the observed area to the surrounding annuli.
The four radial distance ranges $\Delta$ were 2.5--4.25, 4.5--7,
7.25--10, and 10.25--15 heliographic degrees. Only $\tau_{\rm diff}^{\rm oi}$
and $\tau_{\rm mean}$ defined in Sec. 2.3 were used in this case.

 For inversion of these data, we have adopted a four-layer discrete model
described by Kosovichev (1996). The layers which all are 16 Mm thick have
been divided into $72\times 52$ rectangular blocks of approximately the
same size as the spacing between the data points.

The inversion results have shown that perturbations
of the active regions can be seen in the two upper layers, at 0-32 Mm,
but not deeper. The three components of the flow velocity in the upper layer,
inferred from the data, are illustrated in Fig. 11b.
The results show strong downflows of $\approx 1$ km/s around the sunspots,
and some much weaker isolated upflows
at the boundaries of the active latitudes.
The spatial resolution of the data is insufficient to study
the structure and dynamics of individual sunspots.
 However, it is interesting to note that the
peak of the downdraft velocity is observed between the spots in the bipolar
group NOAA 6432 located at latitude $-10^\circ$ and longitude $20^\circ$.
This group results in two separate strong negative sound-speed perturbations
in the top layer (0-16 Mm), which merge into a single perturbation in the second
layer (16-32 Mm).
Figure 12 (top) shows a vertical cut through this active region at a
fixed latitude. The grey-scale map represents the sound speed structure,
and the arrows - the projection of the velocity field into the longitude-depth
plane.
The convergent character of the flow is evident. This active region was only
about 5 days old. The other young active regions in the observed area seem
to show a similar pattern. Contrary, the much older decaying region NOAA 6427
(Fig.12, bottom) shows reduced sound speed and diverging upflow. Another
active region, NOAA 6437, with similar characteristics disappeared a day
after it was observed. It is also interesting to note that
in the flow map (Fig. 11b), the largest flow velocities are observed at the
active latitudes $\pm 10^\circ$ and  $\pm 22^\circ$. Some areas show strong
flows without any apparent activity. One could speculate, that at least,
some of them could be the places
where new magnetic structures emerged later; e.g. the area of strong converging
flow seen in the two upper layers at latitude -8$^\circ$ and longitude
$10^\circ$ was close to the place where a new large group of sunspots,
NOAA 6440, was born two days after.

\section{Conclusion}

Time-distance helioseismology
provides unique information about 3-D structures and flows associated
with magnetic field and turbulent convection. This technique is at the
very beginning of its development.
In this paper, we have presented some basic principles of the
 technique and some initial results of inversion of the travel-time data
obtained from the MDI experiment onboard SOHO (Scherrer, {\it et al.},
1995), and from the geographical South Pole (Jefferies, {\it et al.}, 1994).
The initial results reveal the internal structure of supergranulation
flows, indicating that the divergent outflow in supergranules is only
2--3 Mm deep. The time-distance results also  reveal large-scale
subsurface sound-speed structures and flows related to the active regions.
Also, there is evidence of different flow patterns and
sound-speed structures in the developing and decaying regions.

Obviously, continuous long-term high-resolution observations are necessary
to study the dynamics of solar convection in the quiet and active regions
in detail.

\vspace{3ex} \footnotesize \noindent
{\bf Acknowledgements.}
We are grateful to Danmarks Grundforskningsfonden and TAC for sponsoring
the workshop. This research is partly supported by the SOI-MDI NASA contract
NAG5-3077 at Stanford University.




\end{document}